\documentclass[aip,citeautoscript,
 amsmath,amssymb,
 reprint,
]{revtex4-1}

\usepackage{graphicx}
\usepackage{bm}

\usepackage{bbold}

\begin{document}

\title{Faster Exact Exchange in Periodic Systems using Single-precision Arithmetic }

\author{John Vinson}
\email{john.vinson@nist.gov}
\address{Material Measurement Laboratory, National Institute of Standards and Technology, 100 Bureau Drive, Gaithersburg, MD 20899}

\date{\today}

\begin{abstract}

Density-functional theory simplifies many-electron calculations by approximating the exchange and correlation interactions with a one-electron operator that is a functional of the density.  
Hybrid functionals incorporate some amount of exact exchange, improving agreement with measured electronic and structural properties. 
However, calculations with hybrid functionals require substantial computational resources, limiting their use. 
By calculating the exchange interaction of periodic systems with single-precision arithmetic, the computation time is cut nearly in half with a negligible loss in accuracy. 
This improvement makes exact exchange calculations quicker and more feasible, especially for high-throughput calculations. 
Example hybrid density-functional theory calculations of band energies, forces, and x-ray absorption spectra show that this single-precision implementation maintains accuracy with significantly reduced runtime and memory requirements. 

\end{abstract}

\maketitle

\section{Introduction}

Within density-functional theory (DFT), both the exchange and the correlation interactions are simplified and included as potentials that typically depend only on the local electron density (and its derivatives). 
While DFT has proven to be widely successful, standard local or semi-local functionals have a number of shortcomings that limit accuracy and general applicability. 
One particular problem is that within DFT, each electron interacts with the Coulomb potential of the total electron density. 
This means that each electron is repelled by its own charge density, {\it e.g.}, self-interaction error. 
This is mitigated somewhat by including some fraction of the exact exchange interaction (replacing the functional exchange), and a variety of hybrid  or screened-hybrid functionals incorporating a mix of functional and exact exchange have been proposed \cite{RevModPhys.80.3}.
Despite the success of hybrid functionals, their use has been hindered by the significantly higher computational cost of calculating the exact exchange operator compared to calculations using local or semi-local density functionals. 

In this paper I show that the use of single-precision arithmetic substantially reduces the cost of hybrid density functional calculations while maintaining accuracy. 
First, the exchange operator and its scaling with system size are reviewed. 
Then the Adaptively Compressed Exchange operator approximation is introduced \cite{ACE}, and the implementation of single-precision exact exchange is outlined. 
In Sec.~\ref{sec-results} example calculations are shown that compare both electronic and structural properties between the original, double-precision and new, single-precision implementations of the exact exchange: total electronic energies, band energies, lattice constants, bulk moduli, forces, and x-ray absorption spectra. 
The examined systems include both metals and insulators with unit cells ranging from 1 to 192 atoms. 
Finally, a summary and outlook are presented in Sec.~\ref{sec-outro}.

\section{Review}

The exchange operator $V_X$ is straight-forward,  
\begin{equation}
V_X(\mathbf{r},\mathbf{r}',[\{\psi_i\}]) = -\sum_{i=1}^{N_\mathrm{occ}} \frac{ \psi_i^{*}(\mathbf{r}) \psi_i(\mathbf{r}') }{\vert \mathbf{r} - \mathbf{r}' \vert}
\end{equation}
where $V_X$ is an operator in real-space coordinates $\mathbf{r}$ and $\mathbf{r}'$, and is a functional of the $N_\mathrm{occ}$ occupied electron orbitals $\psi_i$, requiring self-consistency. (In this case self-consistency requires that the eigenfunctions of a Hamiltonian including $V_X$ are the same as those that were used to construct $V_X$.)  
The evaluation of the exact exchange operator, especially in periodic systems, is time consuming. 
This is due in part to the large size of the basis. The number of plane waves is much larger than the number of occupied electron orbitals, $N_G \gg N_\mathrm{occ}$.
Furthermore, the exchange operator must be calculated repeatedly until self-consistency is reached.

A number of methods to reduce the computational expense of evaluating exact exchange in periodic systems have been proposed based on localizing the electron orbitals. 
Localization techniques reduce the spatial extent of individual orbitals in Eq.~1 \cite{PhysRevB.79.085102,doi:10.1021/ct3007088,doi:10.1021/ct500985f}, and are aided by the exponential localization of occupied orbitals in insulating systems \cite{RevModPhys.84.1419, PhysRevLett.98.046402}. 
In large systems these techniques can lead to linear scaling in evaluating the exchange integral, but the localization procedure itself can be costly and its applicability is limited to systems with a band gap.

In non-periodic or molecular systems, exact exchange is much less costly, and Hartree-Fock or post-Hartree-Fock calculations are common-place. 
In simulations of molecular systems, localized basis sets such as Gaussian-type orbitals are typically used.
Not only do these basis functions have finite extent, but the total number of basis functions $N_B$ is much smaller,  $N_G \gg N_B > N_\mathrm{occ}$.
Substantial work has been done to reduce the computational cost of four-center integrals like the exchange. 
The use of reduced-precision has been investigated in quantum chemistry calculations including M{\o}ller-Plesset perturbation theory \cite{doi:10.1021/ct100533u} and coupled-cluster calculations \cite{doi:10.1021/acs.jctc.8b00321}. In both cases it was found to be sufficiently accurate. 
Finite-basis set codes have also shown the utility of reducing the precision for storing pre-computed four-center integrals \cite{doi:10.1063/1.2931945,doi:10.1021/jp908836z}.  
Within the context of periodic systems, single-precision arithmetic can be used to calculate the electron self-energy within the {\sc abinit} code \cite{Gonze2020}. 
However, for periodic, plane-wave approaches reduced precision is rare.

For either finite or periodic systems, with or without a band gap, the Adaptively Compressed Exchange (ACE) operator method reduces the computational cost of exact exchange without compromising accuracy \cite{ACE}. 
The single-precision implementation of exact exchange is built on top of the ACE method. 
A brief overview is presented here.

In the ACE method, the exchange operator $V_X[\{\psi_i\}]$ is still applied to each orbital to create a set 
\begin{equation}
\label{eq-W}
W_j(\mathbf{r}) = [V_X(\mathbf{r},\mathbf{r}',[\{\psi_i\}])\psi_j(\mathbf{r}')](\mathbf{r}) 
\end{equation}
 As noted in Ref.~\onlinecite{ACE}, the construction of $W$ requires $N_\mathrm{occ} N_e$ solutions to a Poisson-like problem, where $N_e \ge N_\mathrm{occ}$ is the number of orbitals being solved for. 
Using fast Fourier transforms to solve the Poisson problem gives a computational cost that scales as $N_\mathrm{occ}N_e N_G \log N_G$.

From $W$, overlaps are taken,
\begin{equation}
M_{jk} = \int d\mathbf{r} \, \psi^{*}_j(\mathbf{r}) W_k(\mathbf{r})
\label{eq-m}
\end{equation}
where $M$ has the dimension $N_e$, and this scales as $N_e^2 N_G$. After a Cholesky factorization $-M = LL^\dagger$, the ACE approximation to the exchange operator is given by 
\begin{eqnarray}
 \xi_i(\mathbf{r}) &= W_j(\mathbf{r})(L^{\dagger})_{ji} \label{eq-xi} \\
V_X^\mathrm{ACE}(\mathbf{r},\mathbf{r}') &= - \sum_{i=1}^{N_e} \xi_i(\mathbf{r}) \xi_i(\mathbf{r}') 
\label{eq-vexxace}
\end{eqnarray}
Two nested self-consistent loop are used to converge the electron orbitals. In the inner loop $V_X^\mathrm{ACE}$ is held constant and only updated once the inner loop has reach self-consistency. 
This greatly reduces the number of times the exchange operator needs to be constructed, but, despite the improvement of the ACE method, including exact exchange still adds significant cost to the calculation. 
The evaluation of Eq.~\ref{eq-W}  scales cubically with system size. Additionally, significant memory is required to store both $W$ and $\xi$.

Here single-precision calculations of the exact exchange for plane wave basis set calculations are introduced, building on top of the ACE method. 
Both memory usage and calculation time are reduced by using single-precision arithmetic to construct and store $W$ in Eq.~\ref{eq-W} and to store $\xi$ in Eqs.~\ref{eq-xi} and \ref{eq-vexxace}. 
In Eq.~\ref{eq-m} the double-precision electron orbitals are used, and the matrices $M$ and $L$ are built in double-precision. 
Mixed-precision matrix operations have not been used. 
Instead, matrices are temporarily converted between precision, {\it e.g.}, $W$ to double in  Eq.~\ref{eq-m} and $L^\dagger$ to single in Eq.~\ref{eq-xi}. 
The conversions were not found to be a significant burden, and mixed-precision linear algebra libraries are not widely available. 
This single-precision approximation reduces the runtime and maximum memory load by almost a factor of 2, with little to no loss of accuracy compared against the double-precision version.

I modified version 6.5 of the {\sc Quantum ESPRESSO} code \cite{espresso2,espresso1}. 
Taking advantage of the iterative nature of the ACE method, an option was also implemented to fall back to the double precision routines after the estimated relative error in the exact exchange energy is below some threshold.  
This allows the speed savings to be realized for the first several iterations, while still producing results at full accuracy. 
This functionality might be necessary for high-precision calculations of large unit cells, {\it e.g.}, for forces. 
However, the error from using single-precision is small, and double-precision calculations of the exact exchange are unnecessary for the examples shown here. 
In what follows I will refer to results and timings of the unmodified, fully double-precision code (DP) and the modified code that uses single-precision for the exact exchange (SP).
To show the success of this single-precision implementation I compare SP and DP calculations for both insulating and metallic systems and both large and small cell sizes.

\section{Results}
\label{sec-results}

As an initial assessment, three crystalline materials are examined: Si, rocksalt ZnS, and Cu.
For all calculations additional convergence parameters are listed in the appendix. 
Pseudopotentials were taken from the PseudoDojo collection \cite{pspdojo1,pspdojo0} with non-linear core corrections removed and generated using the {\sc oncvpsp} code \cite{PhysRevB.88.085117,oncvp}. 
To converge the calculations plane-wave cut-offs and {\it k}-point sampling were chosen to ensure a target total energy convergence of $1.4\times10^{-4}$~Ry per atom (2~meV per atom), and each self-consistent calculation was set to a tolerance of $5\times10^{-8}$~Ry per atom.
For silicon, the lattice constants are taken from experiment \cite{Wyckoff}, while for both ZnS and Cu the structures were relaxed. 
For all three materials the HSE exchange-correlation potential was used \cite{hse03,hse06}. 
The HSE functional is a screened hybrid, where exact exchange in Eq.~1 is reduced by a range-dependent function. 
While screened hybrids tend to converge with fewer {\it k}-points than hybrid functionals, the computational cost is otherwise the same. 
The runtime, total energy, and root-mean-square (RMS) band energies between the two methods of running the exact exchange for each crystalline material are shown in Table~\ref{timingTable}. 
The deviations between the double-precision and single-precision runs are negligible, but the time and memory savings of single precision are substantial.

\begin{table}
\begin{tabular}{ c | c | c | c | c | c | c }
 & Prec.\ & Time & E$_\mathrm{tot}$ & RMS & $\Delta_\mathit{max}$ & Mem.\  \\
 & & (s) & (Ry) & (eV) & (eV) & (\sc{gb}) \\
 \hline
  Si  & DP  & \, 808  & $-16.8099$ &-  & - & 24.8  \\
       & SP  & \, 292 &  $-16.8099 $ & 2.1$\times10^{-4}$ & 1.7$\times10^{-3}$ & 13.3 \\
   \hline
 ZnS   & DP   & 5424  & $-684.937$ &- & - &  45.3 \\
  ({\it rs}) & SP & 2724 & $-684.937$ & 2.5$\times10^{-6}$ & 7.4$\times10^{-6}$& 23.6 \\
 \hline
 Cu  & DP  &  1280 & $-376.314$ & - & - & 49.0\\
   & SP &  \, 765 & $-376.314$ & 1.6$\times10^{-3}$  & 6.1$\times10^{-2}$ & 25.8\\
   \hline
\end{tabular}
\caption{ The runtime, total energies, RMS errors, maximum difference, and approximate peak memory usage for calculating the exact exchange using double and single-precision.
The RMS error of using single precision is calculated using 8 total bands for Si (4 valence and 4 conduction), 21 for ZnS (8 semi-core, 9 valence, and 4 conduction), and 13 for copper (including 4 semi-core).
The maximum absolute difference $\Delta_\mathit{max}$ is determined over the same range of bands. 
}
\label{timingTable}
\end{table}

In the case of copper, the largest deviation in electron eigenvalues is in the conduction bands. Restricting the error analysis to the lowest 11 bands reduces the maximum deviation and root mean square difference (RMS) to less than a meV. Conversely, neglecting the semi-core orbitals has no effect on the maximum and a negligible effect on the RMS. The less favorable timing for the copper is primarily due to the single-precision run requiring 7 iterations to converge the exact exchange energy, while the double-precision run used only 6.

Next, I show the effect of the single-precision exact exchange on calculated structural properties, starting with the equation of state for Cu and  several phases of ZnS. 
For each, the energy as a function of volume was fit to the 3rd order Birch–Murnaghan equation using 9 volumes from approximately 94~\% to 108~\% of the equilibrium volume \cite{PhysRev.71.809}. 
These results are summarized in Table~\ref{EOStable}.
While phase diagrams are a useful output of high-throughput studies, their utility is hampered by shortcomings of standard density functionals. 	
Calculated lattice parameters are often accurate to a few percentage points, but relative stability between phases can be mis-ordered. 
These shortcomings can be partially mitigated using hybrid density functionals. 
The SP calculations for both metallic copper and the three insulating phases of ZnS are able to reproduce the structural parameters to better than 0.05~\%, exceeding the predictive accuracy of the DFT calculations.

\begin{table}[]
\begin{tabular}{ c | c | c | c | c | c | c}
 & Prec.\ & {\it a}& {\it c}  & B$_0$  & B$'_0$ & $\Delta$E$_{tot}$ \\
 & & (a.u.) & (a.u.) & (GPa.) & & (meV) \\
 \hline
 Cu  & DP  & \,  3.423 & - & 132.8 & 5.244 & - \\
    & SP & \, 3.423 & - & 132.8 & 5.243 & - \\
   \hline
 ZnS   & DP   & 10.281 & - & 72.29 & 4.374 & \, \, 0.0  \\ 
  ({\it zb}) & SP & 10.281 & - & 72.29 & 4.373 & \, \, 0.1 \\ 
      \hline
 ZnS   & DP   &\, 7.244 & 11.867 & 74.88 & 4.261 & 160.9 \\
  ({\it w}) & SP & \, 7.244 & 11.867 &74.88 & 4.288 & 160.9 \\
      \hline
ZnS   & DP   & \, 9.614  & -& 92.96 & 4.408 & 699.1 \\ 
  ({\it rs}) & SP & \, 9.614& - & 92.98 & 4.408 & 699.2 \\ 
        \hline
\end{tabular}
\caption{ Calculated lattice vectors ({\it a, c}), bulk modulus (B$_0$), and the first derivative of the bulk modulus with respect to pressure (B$'_0$) for copper and ZnS [zincblende ({\it zb}), wurtzite ({\it w}), and rocksalt ({\it rs})]. For the polymorphs of ZnS, the energy per formula unit ($\Delta$E$_{tot}$) is listed relative to the DP calculation of the zincblende phase. 
}
\label{EOStable}
\end{table}

Switching from crystalline systems to liquid water, the effect of reduced precision on the calculation of atomic forces will now be evaluated. 
First-principles simulations of liquid water and aqueous systems are hampered by the requirements of large cells to incorporate density fluctuations and the need to include nuclear quantum effects due to the importance of the hydrogen motion. 
Recently, this has been addressed by training machine-learning models to simulate the atomic forces with DFT-level accuracy \cite{PhysRevLett.120.143001}. 
These models can be trained on moderately sized simulations, and then the trained model can be run with significantly lower computation cost, allowing longer simulation times and larger cells. 

\begin{figure}
\begin{centering}
\includegraphics[height=3.1in,trim=120 35 120 60,angle=270]{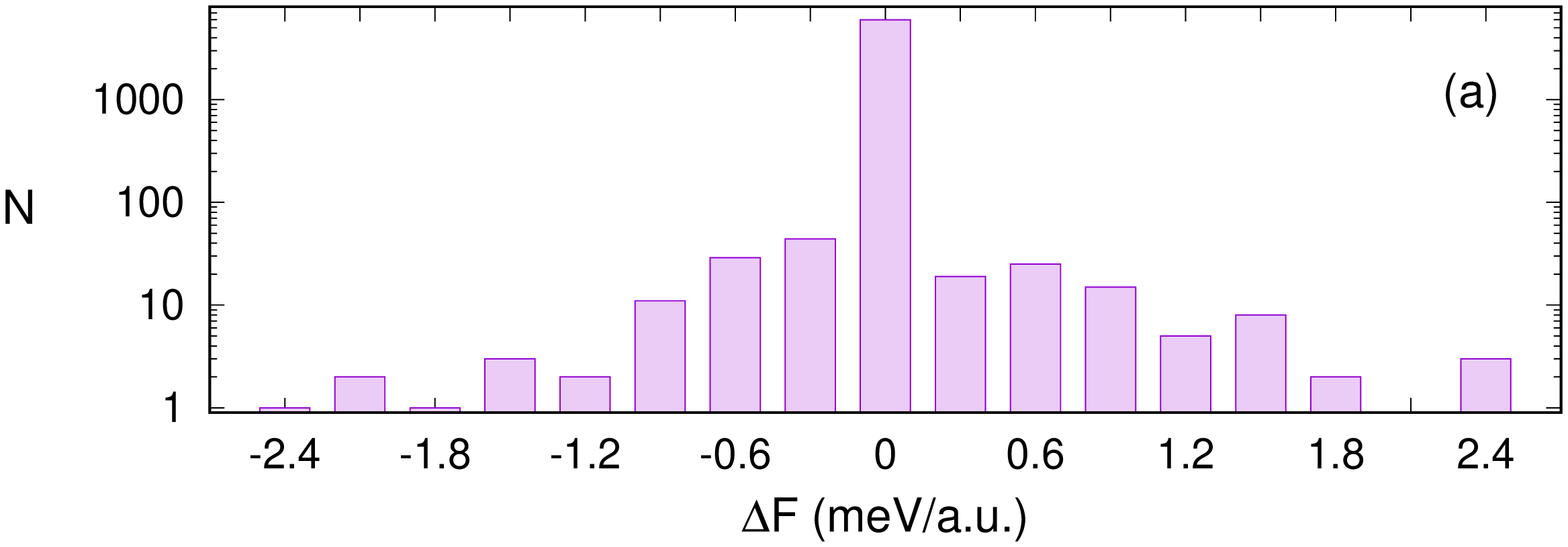} \\
\includegraphics[height=3.1in,trim=120 35 120 60,angle=270]{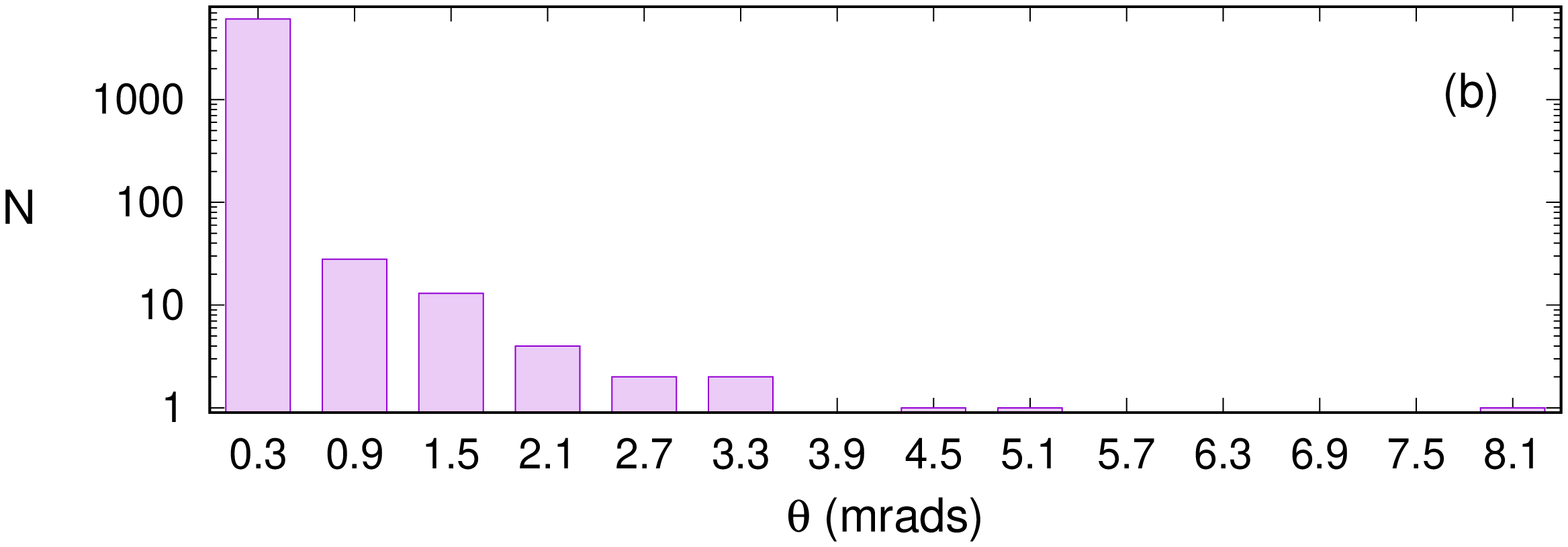}
\caption{ (a) The difference in the magnitudes of the forces between the DP and SP calculations binned into histograms 0.3~meV/a.u.\ wide. (b) The angle between the DP and SP force vectors binned into histograms 0.6~mrads wide. Note the logscale for both. }
\label{waterForcePlot}
\end{centering}
\end{figure}

Snapshots were generated by the Deep Potential Molecular Dynamics model \cite{PhysRevLett.120.143001,CiCP-23-629,doi:10.1080/00268976.2019.1652366}. 
This model was trained using the PBE0 functional \cite{PBE0} and Tkatchenko-Scheffler approximation to the van der Waals interactions \cite{PhysRevLett.102.073005}. 
The calculations here use both. 
The snapshots were generated using path-integral molecular dynamics with 8 beads and NPT conditions: 64 molecules, 300~K, and 100~kPa.\ (1~bar) \cite{PrivateComm}. 
For a DP (SP) run the total memory required per snapshot is 33~GB (18~GB), and the runtime is approximately 169~min.\ (81~min.). 
The forces on each atom are compared between the double-precision and single-precision exact exchange using 4 snapshots spaced more than 1.2~ps apart across all 8 beads (for a total of 6144 forces).  
The difference between the two methods is negligible with a RMS of only 0.142~meV/a.u. and a maximum discrepancy of 2.56~meV/a.u., while the average magnitude of the forces was 800 times larger at 0.12~eV/a.u. 
For comparison, the differences in the forces with and without exact exchange (PBE0 vs.\ PBE) are substantial with an RMS of over 0.25~eV/a.u.
Fig.~\ref{waterForcePlot}(a) shows the distribution of the deviations in the magnitude of the forces between the double-precision and single-precision calculations: for each atom $i$, $\Delta F_i = \| F_i^{\mathrm{DP}} \| - \| F_i^{\mathrm{SP}} \|$. 
In Fig.~\ref{waterForcePlot}(b) the angle between the force vectors is shown: $\theta_i = \mathrm{cos}^{-1} [F_i^{\mathrm{DP}} \cdot F_i^{\mathrm{SP}} /( \| F_i^{\mathrm{DP}} \| \| F_i^{\mathrm{SP}} \| ) ] $.
The average deviation in the angle is less than 19~$\mu$rads and the RMS is 193~$\mu$rads. 

In the case of insulating systems like water, localization techniques such as Wannier projection yield substantial improvements to the computational time needed for exact exchange calculations \cite{doi:10.1021/acs.jctc.9b01167,PhysRevB.79.085102}. 
For metallic systems, localization is less effective, but, as shown previously for copper, there is little change in accuracy when using single precision on gapless systems. 
The SP implementation was tested for moderately-sized metallic systems by investigating dopant sites in AB$_2$O$_4$-type spinel lithium titanate.
 This material has been suggested as a possible lithium battery anode material, though its poor electrical conductivity has lead to attempts to add dopants \cite{doi:10.1021/cm703650c,C3TA11590A}. 
Previous work has shown that Mn dopants likely reside on Li sites through a combination of first-principles modeling and x-ray absorption measurements \cite{PhysRevMaterials.2.125403}. 
A small portion of the analysis of Ref.~\onlinecite{PhysRevMaterials.2.125403} is reproduced here. 
The relative energies of substituting Mn on various Li sites in Li$_7$Ti$_{10}$MnO$_{24}$ is calculated using both HSE and the PBE-sol functionals \cite{PhysRevLett.100.136406}, the latter including simplified Hubbard-$U$ corrections applied to the Ti and Mn {\it d}-orbitals  \cite{PhysRevB.71.035105}. 
As shown in Table~\ref{MnLTO} the relative binding of the single-precision and double-precision HSE are almost the same.
The PBE-sol calculations show that Mn doping the Li-8a$_2$ site is about 0.1~eV per formula unit more stable than doping the Li-8a$_1$ site.
In the HSE calculations, these two sites are degenerate. 
However, the Li-8a$_3$ is still found to be the most stable with both functionals. 

\begin{table}[]
\begin{tabular}{ c | c | c | c | c }
  Site & PBE-sol & HSE-DP & HSE-SP & $\Delta_\mathrm{HSE}$  \\
 \hline
 Li-8a$_3$ \, & - & - & - & $-$0.001 \\
  Li-8a$_2$ \, & 0.104 & 0.176  & 0.175 & \; 0.000 \\
  Li-8a$_1$ \, & 0.219 & 0.173 & 0.173 & \; 0.000 \\
  Li-16d$_2$ & 0.313 & 0.276 & 0.275 & \; 0.000 \\
 Li-16d$_1$ & 0.509 &0.376 & 0.375 & \; 0.000 \\
\end{tabular}
\caption{ Calculated energy penalties in eV for substituting Mn onto various Li sites with respect to the energy of the Li-8a$_3$ site which is found to be the most favorable. Site labels are taken from Ref.~\onlinecite{PhysRevMaterials.2.125403}.  The absolute energy difference between the DP and SP HSE runs, $\Delta_\mathrm{HSE}$, is found to be nearly 0~eV for each structure. 
}
\label{MnLTO}
\end{table}

\begin{figure}
\begin{centering}
\includegraphics[height=3.1in,trim=0 35 15 70,angle=270]{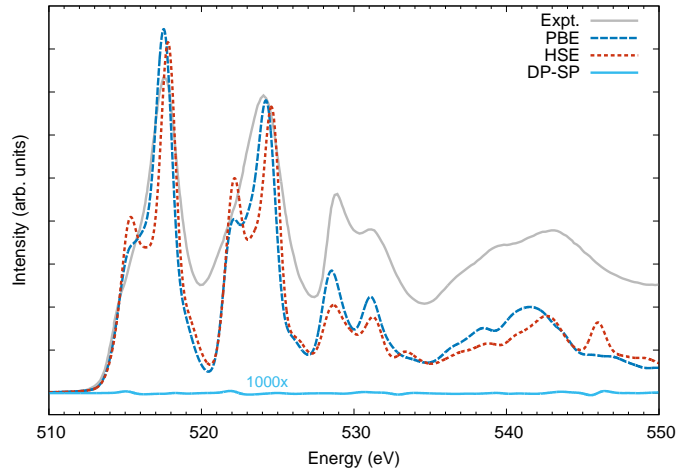}
\caption{ Calculated V L and O K edge x-ray absorption of insulating VO$_2$ compared with experiment \cite{KUMAR2020106335}. 
Separate calculations were carried out using PBE (blue, dashed) or HSE (red, dotted) functionals for the basis states for the Bethe-Salpeter equation. Along the bottom, the solid cyan line shows the difference between the DP and SP HSE calculated spectra, multiplied by a factor of 1000.    }
\label{xas}
\end{centering}
\end{figure}

So far the SP results have been shown to be accurate for the electron energies (both band energies and the total energy) as well as density response (equation of state and forces). 
As a final test I consider the effects of the precision of the exact exchange on x-ray absorption calculations. 
In x-ray absorption, the electron orbitals are probed directly in the form of the transition matrix elements where the electron-photon operator promotes an electron from a core level into the conduction band. 
Using the {\sc ocean} code \cite{ocean0,ocean1}, the x-ray absorption spectra of the M$_1$ insulating phase of VO$_2$ are calculated, looking at  the O K and V L$_{23}$ edges.
The lattice constants and atomic positions were taken from experiment \cite{VO2}. 
The {\sc ocean} calculation uses DFT electron orbitals as the basis for solving the Bethe-Salpeter equation. 
The measured data are compared to calculations using the PBE functional with both SP and DP HSE calculations in Fig.~\ref{xas}. 
All three calculations use the same screened core-hole potential, taken from the PBE calculation. 
The difference between the DP and SP calculations is negligible. In the plot the difference curve is multiplied by a factor of 1000 to make any differences visible. 

Within DFT, local or semi-local exchange-correlation potentials do not correctly describe the insulating phases of VO$_2$. 
Several attempts to rectify this have been attempted, such as including exact exchange \cite{PhysRevLett.107.016401,PhysRevB.95.125105} or self-consistent self-energy calculations \cite{PhysRevLett.99.266402}. 
As criteria, structure and the band gap only reflect the occupied states and lowest-lying conduction band. 
Electronic excitations such as optical or x-ray spectra provide the additional comparison of how hybrid functionals modify the unoccupied states of material. 
These calculations come with an increase in computational cost since they require also determining the unoccupied orbitals.  
Optical excitations in VO$_2$ were calculated using self-consistent self-energy showing a dramatic change in the absorption spectra \cite{PhysRevB.91.195137}. 
However, x-ray absorption calculations of the V L$_{23}$ edge using a similar method found, despite the opening of a gap in the ground state, little effect on the calculated spectra \cite{PhysRevB.86.195135}. 

Here clear differences are evident between the BSE x-ray absorption using PBE versus HSE.  
In VO$_2$, the unoccupied V{\it d} orbitals are split by the symmetry depending on their overlap with the oxygen orbitals \cite{PhysRevB.43.7263}. 
The HSE decreases the occupancy of the lower-energy orbitals, resulting in a stronger transition near 515~eV at the V L$_3$ (522~eV at the L$_2$). 
There is also an increase in the crystal field splitting from the PBE to the HSE calculation, apparent in the shift of the peaks near 517.5~eV and 524.5~eV to higher energy.
The features from 527~eV to 533~eV also reflect the structure of the V{\it d} orbitals, but these are transitions from the O 1{\it s} to {\it p}-like orbitals that have hybridized with the V{\it d} states. 
The HSE calculation gives reduced intensity in this region, indicating less hybridization between the O{\it p} and V{\it d} orbitals. 
In general, the HSE calculation shows worse agreement with the measured x-ray absorption. 
It was suggested in Ref.~\onlinecite{PhysRevB.95.125105} that the vanadium oxides require a reduced percentage of exact exchange, but a full investigation of the exchange dependence of the x-ray spectra is beyond the scope of this work.

\section{Summary}
\label{sec-outro}

I have shown that the computational cost of hybrid DFT calculations, both total run time and required memory, can be nearly cut in half through the use of single-precision arithmetic. 
Any errors from this approximation are below the expected accuracy of the calculations themselves.  
This improvement does not require insulating systems nor localization methods, and is easily adaptable to any plane-wave DFT method. 
While the current implementation requires norm-conserving pseudopotentials and collinear spin, removing these limitations should be straightforward. 
For small to moderate unit cells, the single-precision method introduced here is a substantial improvement for hybrid DFT calculations. 

Part of the success here is due to the relatively small contribution of the exact exchange to the total energy of the electrons. 
For the systems investigated here, the ratio of the Fock energy to total energy ranges from 2\% in Cu to 6\% in liquid water. 
On the other hand the differences in Fock energies between SP and DP calculations are below 1 part in $10^{-5}$ (compared to the relative precision limit of single-precision numbers of approximately $10^{-7}$). 
Taken together, the relative error in the total energy of using the SP exact exchange is around $10^{-7}$ or better. 
This type of reduced precision approach could also be used for evaluating other parts of the DFT Hamiltonian, such as the higher-order semi-local terms for meta-GGA functionals. 

The fundamental system-size scaling of hybrid DFT is unchanged by the reduced-precision implementation. 
For large systems, some other method or approximation, such as the localization techniques mentioned previously, is necessary to make the calculation feasible. 
The use of mixed or reduced precision may also be applicable to these methods and should be investigated.


\appendix
\section{Convergence parameters}

The water calculations were carried out using a single {\it k}-point and the $\Gamma$-point specific routines within {\sc Quantum ESPRESSO}. 
For the water and ZnS systems only the occupied bands were included in the calculation, while for Si, rocksalt ZnS, and Cu the number of bands is specified in Table I. 
In LTO, 200 bands were used for 300 valence electrons. 
The ions in the LTO calculation were relaxed to reduce the forces below $10^{-3}$~Ry/a.u.
For VO$_2$ 120 bands were calculated (100 valence electrons), but the number of ACE projectors was set to only 80. 
For Cu, LTO, and VO$_2$, Fermi-Dirac smearing was used with a broadening parameter of 0.02~Ry. \\

\begin{table}[]
\begin{tabular}{ c | c | c }
 & E$_{\mathrm{cut}}$ (Ry.)  & {\it k}-points \\
 \hline
  Si  & 50  & $8\times8\times8$ \\
  Cu & 110 & $8\times8\times8$  \\
  ZnS ({\it rs}) & 100 & $6\times6\times6$   \\
   ZnS ({\it zb}) & 100 & $4\times4\times4$  \\
     ZnS ({\it w})\, & 100 & $6\times6\times3$  \\ 
     LTO & 100 & $2\times2\times1$   \\ 
     H$_2$O & 90 & $1\times1\times1$    \\
     VO$_2$ & 110 &  $6\times6\times6$   \\
   \hline
\end{tabular}
\caption{ The plane-wave energy cutoff and {\it k}-point sampling for each of the systems studied in this paper. }
\label{Convergence}
\end{table}

\bibliography{exx}

\begin{thebibliography}{44}%
\makeatletter
\providecommand \@ifxundefined [1]{%
 \@ifx{#1\undefined}
}%
\providecommand \@ifnum [1]{%
 \ifnum #1\expandafter \@firstoftwo
 \else \expandafter \@secondoftwo
 \fi
}%
\providecommand \@ifx [1]{%
 \ifx #1\expandafter \@firstoftwo
 \else \expandafter \@secondoftwo
 \fi
}%
\providecommand \natexlab [1]{#1}%
\providecommand \enquote  [1]{``#1''}%
\providecommand \bibnamefont  [1]{#1}%
\providecommand \bibfnamefont [1]{#1}%
\providecommand \citenamefont [1]{#1}%
\providecommand \href@noop [0]{\@secondoftwo}%
\providecommand \href [0]{\begingroup \@sanitize@url \@href}%
\providecommand \@href[1]{\@@startlink{#1}\@@href}%
\providecommand \@@href[1]{\endgroup#1\@@endlink}%
\providecommand \@sanitize@url [0]{\catcode `\\12\catcode `\$12\catcode
  `\&12\catcode `\#12\catcode `\^12\catcode `\_12\catcode `\%12\relax}%
\providecommand \@@startlink[1]{}%
\providecommand \@@endlink[0]{}%
\providecommand \url  [0]{\begingroup\@sanitize@url \@url }%
\providecommand \@url [1]{\endgroup\@href {#1}{\urlprefix }}%
\providecommand \urlprefix  [0]{URL }%
\providecommand \Eprint [0]{\href }%
\providecommand \doibase [0]{https://doi.org/}%
\providecommand \selectlanguage [0]{\@gobble}%
\providecommand \bibinfo  [0]{\@secondoftwo}%
\providecommand \bibfield  [0]{\@secondoftwo}%
\providecommand \translation [1]{[#1]}%
\providecommand \BibitemOpen [0]{}%
\providecommand \bibitemStop [0]{}%
\providecommand \bibitemNoStop [0]{.\EOS\space}%
\providecommand \EOS [0]{\spacefactor3000\relax}%
\providecommand \BibitemShut  [1]{\csname bibitem#1\endcsname}%
\let\auto@bib@innerbib\@empty
\bibitem [{\citenamefont {K\"ummel}\ and\ \citenamefont
  {Kronik}(2008)}]{RevModPhys.80.3}%
  \BibitemOpen
  \bibfield  {author} {\bibinfo {author} {\bibfnamefont {S.}~\bibnamefont
  {K\"ummel}}\ and\ \bibinfo {author} {\bibfnamefont {L.}~\bibnamefont
  {Kronik}},\ }\bibfield  {title} {\enquote {\bibinfo {title}
  {Orbital-dependent density functionals: Theory and applications},}\ }\href
  {https://doi.org/10.1103/RevModPhys.80.3} {\bibfield  {journal} {\bibinfo
  {journal} {Rev. Mod. Phys.}\ }\textbf {\bibinfo {volume} {80}},\ \bibinfo
  {pages} {3--60} (\bibinfo {year} {2008})}\BibitemShut {NoStop}%
\bibitem [{\citenamefont {Lin}(2016)}]{ACE}%
  \BibitemOpen
  \bibfield  {author} {\bibinfo {author} {\bibfnamefont {L.}~\bibnamefont
  {Lin}},\ }\bibfield  {title} {\enquote {\bibinfo {title} {Adaptively
  compressed exchange operator},}\ }\href
  {https://doi.org/10.1021/acs.jctc.6b00092} {\bibfield  {journal} {\bibinfo
  {journal} {Journal of Chemical Theory and Computation}\ }\textbf {\bibinfo
  {volume} {12}},\ \bibinfo {pages} {2242--2249} (\bibinfo {year}
  {2016})}\BibitemShut {NoStop}%
\bibitem [{\citenamefont {Wu}, \citenamefont {Selloni},\ and\ \citenamefont
  {Car}(2009)}]{PhysRevB.79.085102}%
  \BibitemOpen
  \bibfield  {author} {\bibinfo {author} {\bibfnamefont {X.}~\bibnamefont
  {Wu}}, \bibinfo {author} {\bibfnamefont {A.}~\bibnamefont {Selloni}},\ and\
  \bibinfo {author} {\bibfnamefont {R.}~\bibnamefont {Car}},\ }\bibfield
  {title} {\enquote {\bibinfo {title} {Order-$n$ implementation of exact
  exchange in extended insulating systems},}\ }\href
  {https://doi.org/10.1103/PhysRevB.79.085102} {\bibfield  {journal} {\bibinfo
  {journal} {Phys. Rev. B}\ }\textbf {\bibinfo {volume} {79}},\ \bibinfo
  {pages} {085102} (\bibinfo {year} {2009})}\BibitemShut {NoStop}%
\bibitem [{\citenamefont {Gygi}\ and\ \citenamefont
  {Duchemin}(2013)}]{doi:10.1021/ct3007088}%
  \BibitemOpen
  \bibfield  {author} {\bibinfo {author} {\bibfnamefont {F.}~\bibnamefont
  {Gygi}}\ and\ \bibinfo {author} {\bibfnamefont {I.}~\bibnamefont
  {Duchemin}},\ }\bibfield  {title} {\enquote {\bibinfo {title} {Efficient
  computation of hartree–fock exchange using recursive subspace bisection},}\
  }\href {https://doi.org/10.1021/ct3007088} {\bibfield  {journal} {\bibinfo
  {journal} {Journal of Chemical Theory and Computation}\ }\textbf {\bibinfo
  {volume} {9}},\ \bibinfo {pages} {582--587} (\bibinfo {year}
  {2013})}\BibitemShut {NoStop}%
\bibitem [{\citenamefont {Damle}, \citenamefont {Lin},\ and\ \citenamefont
  {Ying}(2015)}]{doi:10.1021/ct500985f}%
  \BibitemOpen
  \bibfield  {author} {\bibinfo {author} {\bibfnamefont {A.}~\bibnamefont
  {Damle}}, \bibinfo {author} {\bibfnamefont {L.}~\bibnamefont {Lin}},\ and\
  \bibinfo {author} {\bibfnamefont {L.}~\bibnamefont {Ying}},\ }\bibfield
  {title} {\enquote {\bibinfo {title} {Compressed representation of kohn–sham
  orbitals via selected columns of the density matrix},}\ }\href
  {https://doi.org/10.1021/ct500985f} {\bibfield  {journal} {\bibinfo
  {journal} {Journal of Chemical Theory and Computation}\ }\textbf {\bibinfo
  {volume} {11}},\ \bibinfo {pages} {1463--1469} (\bibinfo {year}
  {2015})}\BibitemShut {NoStop}%
\bibitem [{\citenamefont {Marzari}\ \emph {et~al.}(2012)\citenamefont
  {Marzari}, \citenamefont {Mostofi}, \citenamefont {Yates}, \citenamefont
  {Souza},\ and\ \citenamefont {Vanderbilt}}]{RevModPhys.84.1419}%
  \BibitemOpen
  \bibfield  {author} {\bibinfo {author} {\bibfnamefont {N.}~\bibnamefont
  {Marzari}}, \bibinfo {author} {\bibfnamefont {A.~A.}\ \bibnamefont
  {Mostofi}}, \bibinfo {author} {\bibfnamefont {J.~R.}\ \bibnamefont {Yates}},
  \bibinfo {author} {\bibfnamefont {I.}~\bibnamefont {Souza}},\ and\ \bibinfo
  {author} {\bibfnamefont {D.}~\bibnamefont {Vanderbilt}},\ }\bibfield  {title}
  {\enquote {\bibinfo {title} {Maximally localized wannier functions: Theory
  and applications},}\ }\href {https://doi.org/10.1103/RevModPhys.84.1419}
  {\bibfield  {journal} {\bibinfo  {journal} {Rev. Mod. Phys.}\ }\textbf
  {\bibinfo {volume} {84}},\ \bibinfo {pages} {1419--1475} (\bibinfo {year}
  {2012})}\BibitemShut {NoStop}%
\bibitem [{\citenamefont {Brouder}\ \emph {et~al.}(2007)\citenamefont
  {Brouder}, \citenamefont {Panati}, \citenamefont {Calandra}, \citenamefont
  {Mourougane},\ and\ \citenamefont {Marzari}}]{PhysRevLett.98.046402}%
  \BibitemOpen
  \bibfield  {author} {\bibinfo {author} {\bibfnamefont {C.}~\bibnamefont
  {Brouder}}, \bibinfo {author} {\bibfnamefont {G.}~\bibnamefont {Panati}},
  \bibinfo {author} {\bibfnamefont {M.}~\bibnamefont {Calandra}}, \bibinfo
  {author} {\bibfnamefont {C.}~\bibnamefont {Mourougane}},\ and\ \bibinfo
  {author} {\bibfnamefont {N.}~\bibnamefont {Marzari}},\ }\bibfield  {title}
  {\enquote {\bibinfo {title} {Exponential localization of wannier functions in
  insulators},}\ }\href {https://doi.org/10.1103/PhysRevLett.98.046402}
  {\bibfield  {journal} {\bibinfo  {journal} {Phys. Rev. Lett.}\ }\textbf
  {\bibinfo {volume} {98}},\ \bibinfo {pages} {046402} (\bibinfo {year}
  {2007})}\BibitemShut {NoStop}%
\bibitem [{\citenamefont {Vysotskiy}\ and\ \citenamefont
  {Cederbaum}(2011)}]{doi:10.1021/ct100533u}%
  \BibitemOpen
  \bibfield  {author} {\bibinfo {author} {\bibfnamefont {V.~P.}\ \bibnamefont
  {Vysotskiy}}\ and\ \bibinfo {author} {\bibfnamefont {L.~S.}\ \bibnamefont
  {Cederbaum}},\ }\bibfield  {title} {\enquote {\bibinfo {title} {Accurate
  quantum chemistry in single precision arithmetic: Correlation energy},}\
  }\href {https://doi.org/10.1021/ct100533u} {\bibfield  {journal} {\bibinfo
  {journal} {Journal of Chemical Theory and Computation}\ }\textbf {\bibinfo
  {volume} {7}},\ \bibinfo {pages} {320--326} (\bibinfo {year}
  {2011})}\BibitemShut {NoStop}%
\bibitem [{\citenamefont {Pokhilko}, \citenamefont {Epifanovsky},\ and\
  \citenamefont {Krylov}(2018)}]{doi:10.1021/acs.jctc.8b00321}%
  \BibitemOpen
  \bibfield  {author} {\bibinfo {author} {\bibfnamefont {P.}~\bibnamefont
  {Pokhilko}}, \bibinfo {author} {\bibfnamefont {E.}~\bibnamefont
  {Epifanovsky}},\ and\ \bibinfo {author} {\bibfnamefont {A.~I.}\ \bibnamefont
  {Krylov}},\ }\bibfield  {title} {\enquote {\bibinfo {title} {Double precision
  is not needed for many-body calculations: Emergent conventional wisdom},}\
  }\href {https://doi.org/10.1021/acs.jctc.8b00321} {\bibfield  {journal}
  {\bibinfo  {journal} {Journal of Chemical Theory and Computation}\ }\textbf
  {\bibinfo {volume} {14}},\ \bibinfo {pages} {4088--4096} (\bibinfo {year}
  {2018})}\BibitemShut {NoStop}%
\bibitem [{\citenamefont {Guidon}\ \emph {et~al.}(2008)\citenamefont {Guidon},
  \citenamefont {Schiffmann}, \citenamefont {Hutter},\ and\ \citenamefont
  {VandeVondele}}]{doi:10.1063/1.2931945}%
  \BibitemOpen
  \bibfield  {author} {\bibinfo {author} {\bibfnamefont {M.}~\bibnamefont
  {Guidon}}, \bibinfo {author} {\bibfnamefont {F.}~\bibnamefont {Schiffmann}},
  \bibinfo {author} {\bibfnamefont {J.}~\bibnamefont {Hutter}},\ and\ \bibinfo
  {author} {\bibfnamefont {J.}~\bibnamefont {VandeVondele}},\ }\bibfield
  {title} {\enquote {\bibinfo {title} {Ab initio molecular dynamics using
  hybrid density functionals},}\ }\href {https://doi.org/10.1063/1.2931945}
  {\bibfield  {journal} {\bibinfo  {journal} {The Journal of Chemical Physics}\
  }\textbf {\bibinfo {volume} {128}},\ \bibinfo {pages} {214104} (\bibinfo
  {year} {2008})}\BibitemShut {NoStop}%
\bibitem [{\citenamefont {Shang}, \citenamefont {Li},\ and\ \citenamefont
  {Yang}(2010)}]{doi:10.1021/jp908836z}%
  \BibitemOpen
  \bibfield  {author} {\bibinfo {author} {\bibfnamefont {H.}~\bibnamefont
  {Shang}}, \bibinfo {author} {\bibfnamefont {Z.}~\bibnamefont {Li}},\ and\
  \bibinfo {author} {\bibfnamefont {J.}~\bibnamefont {Yang}},\ }\bibfield
  {title} {\enquote {\bibinfo {title} {Implementation of exact exchange with
  numerical atomic orbitals},}\ }\href {https://doi.org/10.1021/jp908836z}
  {\bibfield  {journal} {\bibinfo  {journal} {The Journal of Physical Chemistry
  A}\ }\textbf {\bibinfo {volume} {114}},\ \bibinfo {pages} {1039--1043}
  (\bibinfo {year} {2010})},\ \bibinfo {note} {pMID: 20070129}\BibitemShut
  {NoStop}%
\bibitem [{\citenamefont {Gonze}\ \emph {et~al.}(2020)\citenamefont {Gonze},
  \citenamefont {Amadon}, \citenamefont {Antonius}, \citenamefont {Arnardi},
  \citenamefont {Baguet}, \citenamefont {Beuken}, \citenamefont {Bieder},
  \citenamefont {Bottin}, \citenamefont {Bouchet}, \citenamefont {Bousquet},
  \citenamefont {Brouwer}, \citenamefont {Bruneval}, \citenamefont {Brunin},
  \citenamefont {Cavignac}, \citenamefont {Charraud}, \citenamefont {Chen},
  \citenamefont {Côté}, \citenamefont {Cottenier}, \citenamefont {Denier},
  \citenamefont {Geneste}, \citenamefont {Ghosez}, \citenamefont {Giantomassi},
  \citenamefont {Gillet}, \citenamefont {Gingras}, \citenamefont {Hamann},
  \citenamefont {Hautier}, \citenamefont {He}, \citenamefont {Helbig},
  \citenamefont {Holzwarth}, \citenamefont {Jia}, \citenamefont {Jollet},
  \citenamefont {Lafargue-Dit-Hauret}, \citenamefont {Lejaeghere},
  \citenamefont {Marques}, \citenamefont {Martin}, \citenamefont {Martins},
  \citenamefont {Miranda}, \citenamefont {Naccarato}, \citenamefont {Persson},
  \citenamefont {Petretto}, \citenamefont {Planes}, \citenamefont {Pouillon},
  \citenamefont {Prokhorenko}, \citenamefont {Ricci}, \citenamefont
  {Rignanese}, \citenamefont {Romero}, \citenamefont {Schmitt}, \citenamefont
  {Torrent}, \citenamefont {van Setten}, \citenamefont {Troeye}, \citenamefont
  {Verstraete}, \citenamefont {Zérah},\ and\ \citenamefont
  {Zwanziger}}]{Gonze2020}%
  \BibitemOpen
  \bibfield  {author} {\bibinfo {author} {\bibfnamefont {X.}~\bibnamefont
  {Gonze}}, \bibinfo {author} {\bibfnamefont {B.}~\bibnamefont {Amadon}},
  \bibinfo {author} {\bibfnamefont {G.}~\bibnamefont {Antonius}}, \bibinfo
  {author} {\bibfnamefont {F.}~\bibnamefont {Arnardi}}, \bibinfo {author}
  {\bibfnamefont {L.}~\bibnamefont {Baguet}}, \bibinfo {author} {\bibfnamefont
  {J.-M.}\ \bibnamefont {Beuken}}, \bibinfo {author} {\bibfnamefont
  {J.}~\bibnamefont {Bieder}}, \bibinfo {author} {\bibfnamefont
  {F.}~\bibnamefont {Bottin}}, \bibinfo {author} {\bibfnamefont
  {J.}~\bibnamefont {Bouchet}}, \bibinfo {author} {\bibfnamefont
  {E.}~\bibnamefont {Bousquet}}, \bibinfo {author} {\bibfnamefont
  {N.}~\bibnamefont {Brouwer}}, \bibinfo {author} {\bibfnamefont
  {F.}~\bibnamefont {Bruneval}}, \bibinfo {author} {\bibfnamefont
  {G.}~\bibnamefont {Brunin}}, \bibinfo {author} {\bibfnamefont
  {T.}~\bibnamefont {Cavignac}}, \bibinfo {author} {\bibfnamefont {J.-B.}\
  \bibnamefont {Charraud}}, \bibinfo {author} {\bibfnamefont {W.}~\bibnamefont
  {Chen}}, \bibinfo {author} {\bibfnamefont {M.}~\bibnamefont {Côté}},
  \bibinfo {author} {\bibfnamefont {S.}~\bibnamefont {Cottenier}}, \bibinfo
  {author} {\bibfnamefont {J.}~\bibnamefont {Denier}}, \bibinfo {author}
  {\bibfnamefont {G.}~\bibnamefont {Geneste}}, \bibinfo {author} {\bibfnamefont
  {P.}~\bibnamefont {Ghosez}}, \bibinfo {author} {\bibfnamefont
  {M.}~\bibnamefont {Giantomassi}}, \bibinfo {author} {\bibfnamefont
  {Y.}~\bibnamefont {Gillet}}, \bibinfo {author} {\bibfnamefont
  {O.}~\bibnamefont {Gingras}}, \bibinfo {author} {\bibfnamefont {D.~R.}\
  \bibnamefont {Hamann}}, \bibinfo {author} {\bibfnamefont {G.}~\bibnamefont
  {Hautier}}, \bibinfo {author} {\bibfnamefont {X.}~\bibnamefont {He}},
  \bibinfo {author} {\bibfnamefont {N.}~\bibnamefont {Helbig}}, \bibinfo
  {author} {\bibfnamefont {N.}~\bibnamefont {Holzwarth}}, \bibinfo {author}
  {\bibfnamefont {Y.}~\bibnamefont {Jia}}, \bibinfo {author} {\bibfnamefont
  {F.}~\bibnamefont {Jollet}}, \bibinfo {author} {\bibfnamefont
  {W.}~\bibnamefont {Lafargue-Dit-Hauret}}, \bibinfo {author} {\bibfnamefont
  {K.}~\bibnamefont {Lejaeghere}}, \bibinfo {author} {\bibfnamefont {M.~A.~L.}\
  \bibnamefont {Marques}}, \bibinfo {author} {\bibfnamefont {A.}~\bibnamefont
  {Martin}}, \bibinfo {author} {\bibfnamefont {C.}~\bibnamefont {Martins}},
  \bibinfo {author} {\bibfnamefont {H.~P.~C.}\ \bibnamefont {Miranda}},
  \bibinfo {author} {\bibfnamefont {F.}~\bibnamefont {Naccarato}}, \bibinfo
  {author} {\bibfnamefont {K.}~\bibnamefont {Persson}}, \bibinfo {author}
  {\bibfnamefont {G.}~\bibnamefont {Petretto}}, \bibinfo {author}
  {\bibfnamefont {V.}~\bibnamefont {Planes}}, \bibinfo {author} {\bibfnamefont
  {Y.}~\bibnamefont {Pouillon}}, \bibinfo {author} {\bibfnamefont
  {S.}~\bibnamefont {Prokhorenko}}, \bibinfo {author} {\bibfnamefont
  {F.}~\bibnamefont {Ricci}}, \bibinfo {author} {\bibfnamefont {G.-M.}\
  \bibnamefont {Rignanese}}, \bibinfo {author} {\bibfnamefont {A.~H.}\
  \bibnamefont {Romero}}, \bibinfo {author} {\bibfnamefont {M.~M.}\
  \bibnamefont {Schmitt}}, \bibinfo {author} {\bibfnamefont {M.}~\bibnamefont
  {Torrent}}, \bibinfo {author} {\bibfnamefont {M.~J.}\ \bibnamefont {van
  Setten}}, \bibinfo {author} {\bibfnamefont {B.~V.}\ \bibnamefont {Troeye}},
  \bibinfo {author} {\bibfnamefont {M.~J.}\ \bibnamefont {Verstraete}},
  \bibinfo {author} {\bibfnamefont {G.}~\bibnamefont {Zérah}},\ and\ \bibinfo
  {author} {\bibfnamefont {J.~W.}\ \bibnamefont {Zwanziger}},\ }\bibfield
  {title} {\enquote {\bibinfo {title} {The abinit project: Impact, environment
  and recent developments},}\ }\href
  {https://doi.org/10.1016/j.cpc.2019.107042} {\bibfield  {journal} {\bibinfo
  {journal} {Comput. Phys. Commun.}\ }\textbf {\bibinfo {volume} {248}},\
  \bibinfo {pages} {107042} (\bibinfo {year} {2020})}\BibitemShut {NoStop}%
\bibitem [{\citenamefont {Giannozzi}\ \emph {et~al.}(2017)\citenamefont
  {Giannozzi}, \citenamefont {Andreussi}, \citenamefont {Brumme}, \citenamefont
  {Bunau}, \citenamefont {Nardelli}, \citenamefont {Calandra}, \citenamefont
  {Car}, \citenamefont {Cavazzoni}, \citenamefont {Ceresoli}, \citenamefont
  {Cococcioni}, \citenamefont {Colonna}, \citenamefont {Carnimeo},
  \citenamefont {Corso}, \citenamefont {de~Gironcoli}, \citenamefont {Delugas},
  \citenamefont {DiStasio}, \citenamefont {Ferretti}, \citenamefont {Floris},
  \citenamefont {Fratesi}, \citenamefont {Fugallo}, \citenamefont {Gebauer},
  \citenamefont {Gerstmann}, \citenamefont {Giustino}, \citenamefont {Gorni},
  \citenamefont {Jia}, \citenamefont {Kawamura}, \citenamefont {Ko},
  \citenamefont {Kokalj}, \citenamefont {Kü{\c{c}}ükbenli}, \citenamefont
  {Lazzeri}, \citenamefont {Marsili}, \citenamefont {Marzari}, \citenamefont
  {Mauri}, \citenamefont {Nguyen}, \citenamefont {Nguyen}, \citenamefont {de-la
  Roza}, \citenamefont {Paulatto}, \citenamefont {Ponc{\'{e}}}, \citenamefont
  {Rocca}, \citenamefont {Sabatini}, \citenamefont {Santra}, \citenamefont
  {Schlipf}, \citenamefont {Seitsonen}, \citenamefont {Smogunov}, \citenamefont
  {Timrov}, \citenamefont {Thonhauser}, \citenamefont {Umari}, \citenamefont
  {Vast}, \citenamefont {Wu},\ and\ \citenamefont {Baroni}}]{espresso2}%
  \BibitemOpen
  \bibfield  {author} {\bibinfo {author} {\bibfnamefont {P.}~\bibnamefont
  {Giannozzi}}, \bibinfo {author} {\bibfnamefont {O.}~\bibnamefont
  {Andreussi}}, \bibinfo {author} {\bibfnamefont {T.}~\bibnamefont {Brumme}},
  \bibinfo {author} {\bibfnamefont {O.}~\bibnamefont {Bunau}}, \bibinfo
  {author} {\bibfnamefont {M.~B.}\ \bibnamefont {Nardelli}}, \bibinfo {author}
  {\bibfnamefont {M.}~\bibnamefont {Calandra}}, \bibinfo {author}
  {\bibfnamefont {R.}~\bibnamefont {Car}}, \bibinfo {author} {\bibfnamefont
  {C.}~\bibnamefont {Cavazzoni}}, \bibinfo {author} {\bibfnamefont
  {D.}~\bibnamefont {Ceresoli}}, \bibinfo {author} {\bibfnamefont
  {M.}~\bibnamefont {Cococcioni}}, \bibinfo {author} {\bibfnamefont
  {N.}~\bibnamefont {Colonna}}, \bibinfo {author} {\bibfnamefont
  {I.}~\bibnamefont {Carnimeo}}, \bibinfo {author} {\bibfnamefont {A.~D.}\
  \bibnamefont {Corso}}, \bibinfo {author} {\bibfnamefont {S.}~\bibnamefont
  {de~Gironcoli}}, \bibinfo {author} {\bibfnamefont {P.}~\bibnamefont
  {Delugas}}, \bibinfo {author} {\bibfnamefont {R.~A.}\ \bibnamefont
  {DiStasio}}, \bibinfo {author} {\bibfnamefont {A.}~\bibnamefont {Ferretti}},
  \bibinfo {author} {\bibfnamefont {A.}~\bibnamefont {Floris}}, \bibinfo
  {author} {\bibfnamefont {G.}~\bibnamefont {Fratesi}}, \bibinfo {author}
  {\bibfnamefont {G.}~\bibnamefont {Fugallo}}, \bibinfo {author} {\bibfnamefont
  {R.}~\bibnamefont {Gebauer}}, \bibinfo {author} {\bibfnamefont
  {U.}~\bibnamefont {Gerstmann}}, \bibinfo {author} {\bibfnamefont
  {F.}~\bibnamefont {Giustino}}, \bibinfo {author} {\bibfnamefont
  {T.}~\bibnamefont {Gorni}}, \bibinfo {author} {\bibfnamefont
  {J.}~\bibnamefont {Jia}}, \bibinfo {author} {\bibfnamefont {M.}~\bibnamefont
  {Kawamura}}, \bibinfo {author} {\bibfnamefont {H.-Y.}\ \bibnamefont {Ko}},
  \bibinfo {author} {\bibfnamefont {A.}~\bibnamefont {Kokalj}}, \bibinfo
  {author} {\bibfnamefont {E.}~\bibnamefont {Kü{\c{c}}ükbenli}}, \bibinfo
  {author} {\bibfnamefont {M.}~\bibnamefont {Lazzeri}}, \bibinfo {author}
  {\bibfnamefont {M.}~\bibnamefont {Marsili}}, \bibinfo {author} {\bibfnamefont
  {N.}~\bibnamefont {Marzari}}, \bibinfo {author} {\bibfnamefont
  {F.}~\bibnamefont {Mauri}}, \bibinfo {author} {\bibfnamefont {N.~L.}\
  \bibnamefont {Nguyen}}, \bibinfo {author} {\bibfnamefont {H.-V.}\
  \bibnamefont {Nguyen}}, \bibinfo {author} {\bibfnamefont {A.~O.}\
  \bibnamefont {de-la Roza}}, \bibinfo {author} {\bibfnamefont
  {L.}~\bibnamefont {Paulatto}}, \bibinfo {author} {\bibfnamefont
  {S.}~\bibnamefont {Ponc{\'{e}}}}, \bibinfo {author} {\bibfnamefont
  {D.}~\bibnamefont {Rocca}}, \bibinfo {author} {\bibfnamefont
  {R.}~\bibnamefont {Sabatini}}, \bibinfo {author} {\bibfnamefont
  {B.}~\bibnamefont {Santra}}, \bibinfo {author} {\bibfnamefont
  {M.}~\bibnamefont {Schlipf}}, \bibinfo {author} {\bibfnamefont {A.~P.}\
  \bibnamefont {Seitsonen}}, \bibinfo {author} {\bibfnamefont {A.}~\bibnamefont
  {Smogunov}}, \bibinfo {author} {\bibfnamefont {I.}~\bibnamefont {Timrov}},
  \bibinfo {author} {\bibfnamefont {T.}~\bibnamefont {Thonhauser}}, \bibinfo
  {author} {\bibfnamefont {P.}~\bibnamefont {Umari}}, \bibinfo {author}
  {\bibfnamefont {N.}~\bibnamefont {Vast}}, \bibinfo {author} {\bibfnamefont
  {X.}~\bibnamefont {Wu}},\ and\ \bibinfo {author} {\bibfnamefont
  {S.}~\bibnamefont {Baroni}},\ }\bibfield  {title} {\enquote {\bibinfo {title}
  {Advanced capabilities for materials modelling with quantum {ESPRESSO}},}\
  }\href {https://doi.org/10.1088/1361-648x/aa8f79} {\bibfield  {journal}
  {\bibinfo  {journal} {Journal of Physics: Condensed Matter}\ }\textbf
  {\bibinfo {volume} {29}},\ \bibinfo {pages} {465901} (\bibinfo {year}
  {2017})}\BibitemShut {NoStop}%
\bibitem [{\citenamefont {Giannozzi}\ \emph {et~al.}(2009)\citenamefont
  {Giannozzi}, \citenamefont {Baroni}, \citenamefont {Bonini}, \citenamefont
  {Calandra}, \citenamefont {Car}, \citenamefont {Cavazzoni}, \citenamefont
  {Ceresoli}, \citenamefont {Chiarotti}, \citenamefont {Cococcioni},
  \citenamefont {Dabo} \emph {et~al.}}]{espresso1}%
  \BibitemOpen
  \bibfield  {author} {\bibinfo {author} {\bibfnamefont {P.}~\bibnamefont
  {Giannozzi}}, \bibinfo {author} {\bibfnamefont {S.}~\bibnamefont {Baroni}},
  \bibinfo {author} {\bibfnamefont {N.}~\bibnamefont {Bonini}}, \bibinfo
  {author} {\bibfnamefont {M.}~\bibnamefont {Calandra}}, \bibinfo {author}
  {\bibfnamefont {R.}~\bibnamefont {Car}}, \bibinfo {author} {\bibfnamefont
  {C.}~\bibnamefont {Cavazzoni}}, \bibinfo {author} {\bibfnamefont
  {D.}~\bibnamefont {Ceresoli}}, \bibinfo {author} {\bibfnamefont {G.~L.}\
  \bibnamefont {Chiarotti}}, \bibinfo {author} {\bibfnamefont {M.}~\bibnamefont
  {Cococcioni}}, \bibinfo {author} {\bibfnamefont {I.}~\bibnamefont {Dabo}},
  \emph {et~al.},\ }\bibfield  {title} {\enquote {\bibinfo {title} {Quantum
  espresso: a modular and open-source software project for quantum simulations
  of materials},}\ }\href {http://www.quantum-espresso.org} {\bibfield
  {journal} {\bibinfo  {journal} {J. Phys. Condens. Matter}\ }\textbf {\bibinfo
  {volume} {21}},\ \bibinfo {pages} {395502} (\bibinfo {year}
  {2009})}\BibitemShut {NoStop}%
\bibitem [{\citenamefont {van Setten}\ \emph {et~al.}(2018)\citenamefont {van
  Setten}, \citenamefont {Giantomassi}, \citenamefont {Bousquet}, \citenamefont
  {Verstraete}, \citenamefont {Hamann}, \citenamefont {Gonze},\ and\
  \citenamefont {Rignanese}}]{pspdojo1}%
  \BibitemOpen
  \bibfield  {author} {\bibinfo {author} {\bibfnamefont {M.}~\bibnamefont {van
  Setten}}, \bibinfo {author} {\bibfnamefont {M.}~\bibnamefont {Giantomassi}},
  \bibinfo {author} {\bibfnamefont {E.}~\bibnamefont {Bousquet}}, \bibinfo
  {author} {\bibfnamefont {M.}~\bibnamefont {Verstraete}}, \bibinfo {author}
  {\bibfnamefont {D.}~\bibnamefont {Hamann}}, \bibinfo {author} {\bibfnamefont
  {X.}~\bibnamefont {Gonze}},\ and\ \bibinfo {author} {\bibfnamefont {G.-M.}\
  \bibnamefont {Rignanese}},\ }\bibfield  {title} {\enquote {\bibinfo {title}
  {The pseudodojo: Training and grading a 85 element optimized norm-conserving
  pseudopotential table},}\ }\href
  {https://doi.org/https://doi.org/10.1016/j.cpc.2018.01.012} {\bibfield
  {journal} {\bibinfo  {journal} {Computer Physics Communications}\ }\textbf
  {\bibinfo {volume} {226}},\ \bibinfo {pages} {39 -- 54} (\bibinfo {year}
  {2018})}\BibitemShut {NoStop}%
\bibitem [{psp()}]{pspdojo0}%
  \BibitemOpen
  \href@noop {} {}\bibinfo {note} {\texttt{http://www.pseudo-dojo.org} PBE-sol
  Scalar-relativstic {\it v.}~0.4}\BibitemShut {NoStop}%
\bibitem [{\citenamefont {Hamann}(2013)}]{PhysRevB.88.085117}%
  \BibitemOpen
  \bibfield  {author} {\bibinfo {author} {\bibfnamefont {D.~R.}\ \bibnamefont
  {Hamann}},\ }\bibfield  {title} {\enquote {\bibinfo {title} {Optimized
  norm-conserving vanderbilt pseudopotentials},}\ }\href
  {https://doi.org/10.1103/PhysRevB.88.085117} {\bibfield  {journal} {\bibinfo
  {journal} {Phys. Rev. B}\ }\textbf {\bibinfo {volume} {88}},\ \bibinfo
  {pages} {085117} (\bibinfo {year} {2013})}\BibitemShut {NoStop}%
\bibitem [{onc()}]{oncvp}%
  \BibitemOpen
  \href@noop {} {}\bibinfo {note} {The open-source code {\sc oncvpsp} is
  avaiable at \texttt{http://www.mat-simresearch.com} \, {\it
  v.}~3.3.1}\BibitemShut {NoStop}%
\bibitem [{\citenamefont {Wyckoff}(1963)}]{Wyckoff}%
  \BibitemOpen
  \bibfield  {author} {\bibinfo {author} {\bibfnamefont {R.~W.~G.}\
  \bibnamefont {Wyckoff}},\ }\href@noop {} {\emph {\bibinfo {title} {Crystal
  Structures}}},\ \bibinfo {edition} {2nd}\ ed.\ (\bibinfo  {publisher}
  {Interscience Publishers},\ \bibinfo {address} {New York},\ \bibinfo {year}
  {1963})\BibitemShut {NoStop}%
\bibitem [{\citenamefont {Heyd}, \citenamefont {Scuseria},\ and\ \citenamefont
  {Ernzerhof}(2003)}]{hse03}%
  \BibitemOpen
  \bibfield  {author} {\bibinfo {author} {\bibfnamefont {J.}~\bibnamefont
  {Heyd}}, \bibinfo {author} {\bibfnamefont {G.~E.}\ \bibnamefont {Scuseria}},\
  and\ \bibinfo {author} {\bibfnamefont {M.}~\bibnamefont {Ernzerhof}},\
  }\bibfield  {title} {\enquote {\bibinfo {title} {Hybrid functionals based on
  a screened coulomb potential},}\ }\href {https://doi.org/10.1063/1.1564060}
  {\bibfield  {journal} {\bibinfo  {journal} {The Journal of Chemical Physics}\
  }\textbf {\bibinfo {volume} {118}},\ \bibinfo {pages} {8207--8215} (\bibinfo
  {year} {2003})}\BibitemShut {NoStop}%
\bibitem [{\citenamefont {Heyd}, \citenamefont {Scuseria},\ and\ \citenamefont
  {Ernzerhof}(2006)}]{hse06}%
  \BibitemOpen
  \bibfield  {author} {\bibinfo {author} {\bibfnamefont {J.}~\bibnamefont
  {Heyd}}, \bibinfo {author} {\bibfnamefont {G.~E.}\ \bibnamefont {Scuseria}},\
  and\ \bibinfo {author} {\bibfnamefont {M.}~\bibnamefont {Ernzerhof}},\
  }\bibfield  {title} {\enquote {\bibinfo {title} {Erratum: “hybrid
  functionals based on a screened coulomb potential” [j. chem. phys. 118,
  8207 (2003)]},}\ }\href {https://doi.org/10.1063/1.2204597} {\bibfield
  {journal} {\bibinfo  {journal} {The Journal of Chemical Physics}\ }\textbf
  {\bibinfo {volume} {124}},\ \bibinfo {pages} {219906} (\bibinfo {year}
  {2006})}\BibitemShut {NoStop}%
\bibitem [{\citenamefont {Birch}(1947)}]{PhysRev.71.809}%
  \BibitemOpen
  \bibfield  {author} {\bibinfo {author} {\bibfnamefont {F.}~\bibnamefont
  {Birch}},\ }\bibfield  {title} {\enquote {\bibinfo {title} {Finite elastic
  strain of cubic crystals},}\ }\href {https://doi.org/10.1103/PhysRev.71.809}
  {\bibfield  {journal} {\bibinfo  {journal} {Phys. Rev.}\ }\textbf {\bibinfo
  {volume} {71}},\ \bibinfo {pages} {809--824} (\bibinfo {year}
  {1947})}\BibitemShut {NoStop}%
\bibitem [{\citenamefont {Zhang}\ \emph {et~al.}(2018)\citenamefont {Zhang},
  \citenamefont {Han}, \citenamefont {Wang}, \citenamefont {Car},\ and\
  \citenamefont {E}}]{PhysRevLett.120.143001}%
  \BibitemOpen
  \bibfield  {author} {\bibinfo {author} {\bibfnamefont {L.}~\bibnamefont
  {Zhang}}, \bibinfo {author} {\bibfnamefont {J.}~\bibnamefont {Han}}, \bibinfo
  {author} {\bibfnamefont {H.}~\bibnamefont {Wang}}, \bibinfo {author}
  {\bibfnamefont {R.}~\bibnamefont {Car}},\ and\ \bibinfo {author}
  {\bibfnamefont {W.}~\bibnamefont {E}},\ }\bibfield  {title} {\enquote
  {\bibinfo {title} {Deep potential molecular dynamics: A scalable model with
  the accuracy of quantum mechanics},}\ }\href
  {https://doi.org/10.1103/PhysRevLett.120.143001} {\bibfield  {journal}
  {\bibinfo  {journal} {Phys. Rev. Lett.}\ }\textbf {\bibinfo {volume} {120}},\
  \bibinfo {pages} {143001} (\bibinfo {year} {2018})}\BibitemShut {NoStop}%
\bibitem [{CiC(2018)}]{CiCP-23-629}%
  \BibitemOpen
  \bibfield  {title} {\enquote {\bibinfo {title} {Deep potential: A general
  representation of a many-body potential energy surface},}\ }\href
  {https://doi.org/https://doi.org/10.4208/cicp.OA-2017-0213} {\bibfield
  {journal} {\bibinfo  {journal} {Communications in Computational Physics}\
  }\textbf {\bibinfo {volume} {23}},\ \bibinfo {pages} {629--639} (\bibinfo
  {year} {2018})}\BibitemShut {NoStop}%
\bibitem [{\citenamefont {Ko}\ \emph {et~al.}(2019)\citenamefont {Ko},
  \citenamefont {Zhang}, \citenamefont {Santra}, \citenamefont {Wang},
  \citenamefont {E}, \citenamefont {Jr},\ and\ \citenamefont
  {Car}}]{doi:10.1080/00268976.2019.1652366}%
  \BibitemOpen
  \bibfield  {author} {\bibinfo {author} {\bibfnamefont {H.-Y.}\ \bibnamefont
  {Ko}}, \bibinfo {author} {\bibfnamefont {L.}~\bibnamefont {Zhang}}, \bibinfo
  {author} {\bibfnamefont {B.}~\bibnamefont {Santra}}, \bibinfo {author}
  {\bibfnamefont {H.}~\bibnamefont {Wang}}, \bibinfo {author} {\bibfnamefont
  {W.}~\bibnamefont {E}}, \bibinfo {author} {\bibfnamefont {R.~A.~D.}\
  \bibnamefont {Jr}},\ and\ \bibinfo {author} {\bibfnamefont {R.}~\bibnamefont
  {Car}},\ }\bibfield  {title} {\enquote {\bibinfo {title} {Isotope effects in
  liquid water via deep potential molecular dynamics},}\ }\href
  {https://doi.org/10.1080/00268976.2019.1652366} {\bibfield  {journal}
  {\bibinfo  {journal} {Molecular Physics}\ }\textbf {\bibinfo {volume}
  {117}},\ \bibinfo {pages} {3269--3281} (\bibinfo {year} {2019})}\BibitemShut
  {NoStop}%
\bibitem [{\citenamefont {Perdew}, \citenamefont {Ernzerhof},\ and\
  \citenamefont {Burke}(1996)}]{PBE0}%
  \BibitemOpen
  \bibfield  {author} {\bibinfo {author} {\bibfnamefont {J.~P.}\ \bibnamefont
  {Perdew}}, \bibinfo {author} {\bibfnamefont {M.}~\bibnamefont {Ernzerhof}},\
  and\ \bibinfo {author} {\bibfnamefont {K.}~\bibnamefont {Burke}},\ }\bibfield
   {title} {\enquote {\bibinfo {title} {Rationale for mixing exact exchange
  with density functional approximations},}\ }\href
  {https://doi.org/10.1063/1.472933} {\bibfield  {journal} {\bibinfo  {journal}
  {The Journal of Chemical Physics}\ }\textbf {\bibinfo {volume} {105}},\
  \bibinfo {pages} {9982--9985} (\bibinfo {year} {1996})}\BibitemShut {NoStop}%
\bibitem [{\citenamefont {Tkatchenko}\ and\ \citenamefont
  {Scheffler}(2009)}]{PhysRevLett.102.073005}%
  \BibitemOpen
  \bibfield  {author} {\bibinfo {author} {\bibfnamefont {A.}~\bibnamefont
  {Tkatchenko}}\ and\ \bibinfo {author} {\bibfnamefont {M.}~\bibnamefont
  {Scheffler}},\ }\bibfield  {title} {\enquote {\bibinfo {title} {Accurate
  molecular van der waals interactions from ground-state electron density and
  free-atom reference data},}\ }\href
  {https://doi.org/10.1103/PhysRevLett.102.073005} {\bibfield  {journal}
  {\bibinfo  {journal} {Phys. Rev. Lett.}\ }\textbf {\bibinfo {volume} {102}},\
  \bibinfo {pages} {073005} (\bibinfo {year} {2009})}\BibitemShut {NoStop}%
\bibitem [{\citenamefont {Calegari}\ and\ \citenamefont {Car}()}]{PrivateComm}%
  \BibitemOpen
  \bibfield  {author} {\bibinfo {author} {\bibfnamefont {M.}~\bibnamefont
  {Calegari}}\ and\ \bibinfo {author} {\bibfnamefont {R.}~\bibnamefont {Car}},\
  }\href@noop {} {}\bibinfo {note} {Private communication}\BibitemShut
  {NoStop}%
\bibitem [{\citenamefont {Ko}\ \emph {et~al.}(0)\citenamefont {Ko},
  \citenamefont {Jia}, \citenamefont {Santra}, \citenamefont {Wu},
  \citenamefont {Car},\ and\ \citenamefont
  {DiStasio~Jr.}}]{doi:10.1021/acs.jctc.9b01167}%
  \BibitemOpen
  \bibfield  {author} {\bibinfo {author} {\bibfnamefont {H.-Y.}\ \bibnamefont
  {Ko}}, \bibinfo {author} {\bibfnamefont {J.}~\bibnamefont {Jia}}, \bibinfo
  {author} {\bibfnamefont {B.}~\bibnamefont {Santra}}, \bibinfo {author}
  {\bibfnamefont {X.}~\bibnamefont {Wu}}, \bibinfo {author} {\bibfnamefont
  {R.}~\bibnamefont {Car}},\ and\ \bibinfo {author} {\bibfnamefont {R.~A.}\
  \bibnamefont {DiStasio~Jr.}},\ }\bibfield  {title} {\enquote {\bibinfo
  {title} {Enabling large-scale condensed-phase hybrid density functional
  theory based ab initio molecular dynamics. 1. theory, algorithm, and
  performance},}\ }\href {https://doi.org/10.1021/acs.jctc.9b01167} {\bibfield
  {journal} {\bibinfo  {journal} {Journal of Chemical Theory and Computation}\
  }\textbf {\bibinfo {volume} {0}},\ \bibinfo {pages} {null} (\bibinfo {year}
  {0})}\BibitemShut {NoStop}%
\bibitem [{\citenamefont {Capsoni}\ \emph {et~al.}(2008)\citenamefont
  {Capsoni}, \citenamefont {Bini}, \citenamefont {Massarotti}, \citenamefont
  {Mustarelli}, \citenamefont {Chiodelli}, \citenamefont {Azzoni},
  \citenamefont {Mozzati}, \citenamefont {Linati},\ and\ \citenamefont
  {Ferrari}}]{doi:10.1021/cm703650c}%
  \BibitemOpen
  \bibfield  {author} {\bibinfo {author} {\bibfnamefont {D.}~\bibnamefont
  {Capsoni}}, \bibinfo {author} {\bibfnamefont {M.}~\bibnamefont {Bini}},
  \bibinfo {author} {\bibfnamefont {V.}~\bibnamefont {Massarotti}}, \bibinfo
  {author} {\bibfnamefont {P.}~\bibnamefont {Mustarelli}}, \bibinfo {author}
  {\bibfnamefont {G.}~\bibnamefont {Chiodelli}}, \bibinfo {author}
  {\bibfnamefont {C.~B.}\ \bibnamefont {Azzoni}}, \bibinfo {author}
  {\bibfnamefont {M.~C.}\ \bibnamefont {Mozzati}}, \bibinfo {author}
  {\bibfnamefont {L.}~\bibnamefont {Linati}},\ and\ \bibinfo {author}
  {\bibfnamefont {S.}~\bibnamefont {Ferrari}},\ }\bibfield  {title} {\enquote
  {\bibinfo {title} {Cations distribution and valence states in mn-substituted
  li4ti5o12 structure},}\ }\href {https://doi.org/10.1021/cm703650c} {\bibfield
   {journal} {\bibinfo  {journal} {Chemistry of Materials}\ }\textbf {\bibinfo
  {volume} {20}},\ \bibinfo {pages} {4291--4298} (\bibinfo {year}
  {2008})}\BibitemShut {NoStop}%
\bibitem [{\citenamefont {Kaftelen}\ \emph {et~al.}(2013)\citenamefont
  {Kaftelen}, \citenamefont {Tuncer}, \citenamefont {Tu}, \citenamefont {Repp},
  \citenamefont {Göçmez}, \citenamefont {Thomann}, \citenamefont {Weber},\
  and\ \citenamefont {Erdem}}]{C3TA11590A}%
  \BibitemOpen
  \bibfield  {author} {\bibinfo {author} {\bibfnamefont {H.}~\bibnamefont
  {Kaftelen}}, \bibinfo {author} {\bibfnamefont {M.}~\bibnamefont {Tuncer}},
  \bibinfo {author} {\bibfnamefont {S.}~\bibnamefont {Tu}}, \bibinfo {author}
  {\bibfnamefont {S.}~\bibnamefont {Repp}}, \bibinfo {author} {\bibfnamefont
  {H.}~\bibnamefont {Göçmez}}, \bibinfo {author} {\bibfnamefont
  {R.}~\bibnamefont {Thomann}}, \bibinfo {author} {\bibfnamefont
  {S.}~\bibnamefont {Weber}},\ and\ \bibinfo {author} {\bibfnamefont
  {E.}~\bibnamefont {Erdem}},\ }\bibfield  {title} {\enquote {\bibinfo {title}
  {Mn-substituted spinel li4ti5o12 materials studied by multifrequency epr
  spectroscopy},}\ }\href {https://doi.org/10.1039/C3TA11590A} {\bibfield
  {journal} {\bibinfo  {journal} {J. Mater. Chem. A}\ }\textbf {\bibinfo
  {volume} {1}},\ \bibinfo {pages} {9973--9982} (\bibinfo {year}
  {2013})}\BibitemShut {NoStop}%
\bibitem [{\citenamefont {Singh}\ \emph {et~al.}(2018)\citenamefont {Singh},
  \citenamefont {Topsakal}, \citenamefont {Attenkofer}, \citenamefont {Wolf},
  \citenamefont {Leskes}, \citenamefont {Duan}, \citenamefont {Wang},
  \citenamefont {Vinson}, \citenamefont {Lu},\ and\ \citenamefont
  {Frenkel}}]{PhysRevMaterials.2.125403}%
  \BibitemOpen
  \bibfield  {author} {\bibinfo {author} {\bibfnamefont {H.}~\bibnamefont
  {Singh}}, \bibinfo {author} {\bibfnamefont {M.}~\bibnamefont {Topsakal}},
  \bibinfo {author} {\bibfnamefont {K.}~\bibnamefont {Attenkofer}}, \bibinfo
  {author} {\bibfnamefont {T.}~\bibnamefont {Wolf}}, \bibinfo {author}
  {\bibfnamefont {M.}~\bibnamefont {Leskes}}, \bibinfo {author} {\bibfnamefont
  {Y.}~\bibnamefont {Duan}}, \bibinfo {author} {\bibfnamefont {F.}~\bibnamefont
  {Wang}}, \bibinfo {author} {\bibfnamefont {J.}~\bibnamefont {Vinson}},
  \bibinfo {author} {\bibfnamefont {D.}~\bibnamefont {Lu}},\ and\ \bibinfo
  {author} {\bibfnamefont {A.~I.}\ \bibnamefont {Frenkel}},\ }\bibfield
  {title} {\enquote {\bibinfo {title} {Identification of dopant site and its
  effect on electrochemical activity in mn-doped lithium titanate},}\ }\href
  {https://doi.org/10.1103/PhysRevMaterials.2.125403} {\bibfield  {journal}
  {\bibinfo  {journal} {Phys. Rev. Materials}\ }\textbf {\bibinfo {volume}
  {2}},\ \bibinfo {pages} {125403} (\bibinfo {year} {2018})}\BibitemShut
  {NoStop}%
\bibitem [{\citenamefont {Perdew}\ \emph {et~al.}(2008)\citenamefont {Perdew},
  \citenamefont {Ruzsinszky}, \citenamefont {Csonka}, \citenamefont {Vydrov},
  \citenamefont {Scuseria}, \citenamefont {Constantin}, \citenamefont {Zhou},\
  and\ \citenamefont {Burke}}]{PhysRevLett.100.136406}%
  \BibitemOpen
  \bibfield  {author} {\bibinfo {author} {\bibfnamefont {J.~P.}\ \bibnamefont
  {Perdew}}, \bibinfo {author} {\bibfnamefont {A.}~\bibnamefont {Ruzsinszky}},
  \bibinfo {author} {\bibfnamefont {G.~I.}\ \bibnamefont {Csonka}}, \bibinfo
  {author} {\bibfnamefont {O.~A.}\ \bibnamefont {Vydrov}}, \bibinfo {author}
  {\bibfnamefont {G.~E.}\ \bibnamefont {Scuseria}}, \bibinfo {author}
  {\bibfnamefont {L.~A.}\ \bibnamefont {Constantin}}, \bibinfo {author}
  {\bibfnamefont {X.}~\bibnamefont {Zhou}},\ and\ \bibinfo {author}
  {\bibfnamefont {K.}~\bibnamefont {Burke}},\ }\bibfield  {title} {\enquote
  {\bibinfo {title} {Restoring the density-gradient expansion for exchange in
  solids and surfaces},}\ }\href
  {https://doi.org/10.1103/PhysRevLett.100.136406} {\bibfield  {journal}
  {\bibinfo  {journal} {Phys. Rev. Lett.}\ }\textbf {\bibinfo {volume} {100}},\
  \bibinfo {pages} {136406} (\bibinfo {year} {2008})}\BibitemShut {NoStop}%
\bibitem [{\citenamefont {Cococcioni}\ and\ \citenamefont
  {de~Gironcoli}(2005)}]{PhysRevB.71.035105}%
  \BibitemOpen
  \bibfield  {author} {\bibinfo {author} {\bibfnamefont {M.}~\bibnamefont
  {Cococcioni}}\ and\ \bibinfo {author} {\bibfnamefont {S.}~\bibnamefont
  {de~Gironcoli}},\ }\bibfield  {title} {\enquote {\bibinfo {title} {Linear
  response approach to the calculation of the effective interaction parameters
  in the $\mathrm{LDA}+\mathrm{U}$ method},}\ }\href
  {https://doi.org/10.1103/PhysRevB.71.035105} {\bibfield  {journal} {\bibinfo
  {journal} {Phys. Rev. B}\ }\textbf {\bibinfo {volume} {71}},\ \bibinfo
  {pages} {035105} (\bibinfo {year} {2005})}\BibitemShut {NoStop}%
\bibitem [{\citenamefont {Vinson}\ \emph {et~al.}(2011)\citenamefont {Vinson},
  \citenamefont {Rehr}, \citenamefont {Kas},\ and\ \citenamefont
  {Shirley}}]{ocean0}%
  \BibitemOpen
  \bibfield  {author} {\bibinfo {author} {\bibfnamefont {J.}~\bibnamefont
  {Vinson}}, \bibinfo {author} {\bibfnamefont {J.~J.}\ \bibnamefont {Rehr}},
  \bibinfo {author} {\bibfnamefont {J.~J.}\ \bibnamefont {Kas}},\ and\ \bibinfo
  {author} {\bibfnamefont {E.~L.}\ \bibnamefont {Shirley}},\ }\bibfield
  {title} {\enquote {\bibinfo {title} {Bethe-salpeter equation calculations of
  core excitation spectra},}\ }\href
  {https://doi.org/10.1103/PhysRevB.83.115106} {\bibfield  {journal} {\bibinfo
  {journal} {Phys. Rev. B}\ }\textbf {\bibinfo {volume} {83}},\ \bibinfo
  {pages} {115106} (\bibinfo {year} {2011})}\BibitemShut {NoStop}%
\bibitem [{\citenamefont {Gilmore}\ \emph {et~al.}(2015)\citenamefont
  {Gilmore}, \citenamefont {Vinson}, \citenamefont {Shirley}, \citenamefont
  {Prendergast}, \citenamefont {Pemmaraju}, \citenamefont {Kas}, \citenamefont
  {Vila},\ and\ \citenamefont {Rehr}}]{ocean1}%
  \BibitemOpen
  \bibfield  {author} {\bibinfo {author} {\bibfnamefont {K.}~\bibnamefont
  {Gilmore}}, \bibinfo {author} {\bibfnamefont {J.}~\bibnamefont {Vinson}},
  \bibinfo {author} {\bibfnamefont {E.}~\bibnamefont {Shirley}}, \bibinfo
  {author} {\bibfnamefont {D.}~\bibnamefont {Prendergast}}, \bibinfo {author}
  {\bibfnamefont {C.}~\bibnamefont {Pemmaraju}}, \bibinfo {author}
  {\bibfnamefont {J.}~\bibnamefont {Kas}}, \bibinfo {author} {\bibfnamefont
  {F.}~\bibnamefont {Vila}},\ and\ \bibinfo {author} {\bibfnamefont
  {J.}~\bibnamefont {Rehr}},\ }\bibfield  {title} {\enquote {\bibinfo {title}
  {Efficient implementation of core-excitation bethe-salpeter equation
  calculations},}\ }\href
  {https://doi.org/http://dx.doi.org/10.1016/j.cpc.2015.08.014} {\bibfield
  {journal} {\bibinfo  {journal} {Comput. Phys. Comm.}\ }\textbf {\bibinfo
  {volume} {197}},\ \bibinfo {pages} {109 -- 117} (\bibinfo {year}
  {2015})}\BibitemShut {NoStop}%
\bibitem [{\citenamefont {Andresson}(1956)}]{VO2}%
  \BibitemOpen
  \bibfield  {author} {\bibinfo {author} {\bibfnamefont {G.}~\bibnamefont
  {Andresson}},\ }\bibfield  {title} {\enquote {\bibinfo {title} {Studies on
  vanadium oxides ii. the crystal structure of vanadium dioxide},}\ }\href@noop
  {} {\bibfield  {journal} {\bibinfo  {journal} {Acta Chemica Scandinavica}\
  }\textbf {\bibinfo {volume} {10}},\ \bibinfo {pages} {623} (\bibinfo {year}
  {1956})}\BibitemShut {NoStop}%
\bibitem [{\citenamefont {Kumar}\ \emph {et~al.}(2020)\citenamefont {Kumar},
  \citenamefont {Singh}, \citenamefont {Chae}, \citenamefont {Park},\ and\
  \citenamefont {Lee}}]{KUMAR2020106335}%
  \BibitemOpen
  \bibfield  {author} {\bibinfo {author} {\bibfnamefont {M.}~\bibnamefont
  {Kumar}}, \bibinfo {author} {\bibfnamefont {J.~P.}\ \bibnamefont {Singh}},
  \bibinfo {author} {\bibfnamefont {K.~H.}\ \bibnamefont {Chae}}, \bibinfo
  {author} {\bibfnamefont {J.}~\bibnamefont {Park}},\ and\ \bibinfo {author}
  {\bibfnamefont {H.~H.}\ \bibnamefont {Lee}},\ }\bibfield  {title} {\enquote
  {\bibinfo {title} {Annealing effect on phase transition and thermochromic
  properties of vo2 thin films},}\ }\href
  {https://doi.org/https://doi.org/10.1016/j.spmi.2019.106335} {\bibfield
  {journal} {\bibinfo  {journal} {Superlattices and Microstructures}\ }\textbf
  {\bibinfo {volume} {137}},\ \bibinfo {pages} {106335} (\bibinfo {year}
  {2020})}\BibitemShut {NoStop}%
\bibitem [{\citenamefont {Eyert}(2011)}]{PhysRevLett.107.016401}%
  \BibitemOpen
  \bibfield  {author} {\bibinfo {author} {\bibfnamefont {V.}~\bibnamefont
  {Eyert}},\ }\bibfield  {title} {\enquote {\bibinfo {title}
  {${\mathrm{vo}}_{2}$: A novel view from band theory},}\ }\href
  {https://doi.org/10.1103/PhysRevLett.107.016401} {\bibfield  {journal}
  {\bibinfo  {journal} {Phys. Rev. Lett.}\ }\textbf {\bibinfo {volume} {107}},\
  \bibinfo {pages} {016401} (\bibinfo {year} {2011})}\BibitemShut {NoStop}%
\bibitem [{\citenamefont {Xu}\ \emph {et~al.}(2017)\citenamefont {Xu},
  \citenamefont {Shen}, \citenamefont {Hallman}, \citenamefont {Haglund},\ and\
  \citenamefont {Pantelides}}]{PhysRevB.95.125105}%
  \BibitemOpen
  \bibfield  {author} {\bibinfo {author} {\bibfnamefont {S.}~\bibnamefont
  {Xu}}, \bibinfo {author} {\bibfnamefont {X.}~\bibnamefont {Shen}}, \bibinfo
  {author} {\bibfnamefont {K.~A.}\ \bibnamefont {Hallman}}, \bibinfo {author}
  {\bibfnamefont {R.~F.}\ \bibnamefont {Haglund}},\ and\ \bibinfo {author}
  {\bibfnamefont {S.~T.}\ \bibnamefont {Pantelides}},\ }\bibfield  {title}
  {\enquote {\bibinfo {title} {Unified band-theoretic description of
  structural, electronic, and magnetic properties of vanadium dioxide
  phases},}\ }\href {https://doi.org/10.1103/PhysRevB.95.125105} {\bibfield
  {journal} {\bibinfo  {journal} {Phys. Rev. B}\ }\textbf {\bibinfo {volume}
  {95}},\ \bibinfo {pages} {125105} (\bibinfo {year} {2017})}\BibitemShut
  {NoStop}%
\bibitem [{\citenamefont {Gatti}\ \emph {et~al.}(2007)\citenamefont {Gatti},
  \citenamefont {Bruneval}, \citenamefont {Olevano},\ and\ \citenamefont
  {Reining}}]{PhysRevLett.99.266402}%
  \BibitemOpen
  \bibfield  {author} {\bibinfo {author} {\bibfnamefont {M.}~\bibnamefont
  {Gatti}}, \bibinfo {author} {\bibfnamefont {F.}~\bibnamefont {Bruneval}},
  \bibinfo {author} {\bibfnamefont {V.}~\bibnamefont {Olevano}},\ and\ \bibinfo
  {author} {\bibfnamefont {L.}~\bibnamefont {Reining}},\ }\bibfield  {title}
  {\enquote {\bibinfo {title} {Understanding correlations in vanadium dioxide
  from first principles},}\ }\href
  {https://doi.org/10.1103/PhysRevLett.99.266402} {\bibfield  {journal}
  {\bibinfo  {journal} {Phys. Rev. Lett.}\ }\textbf {\bibinfo {volume} {99}},\
  \bibinfo {pages} {266402} (\bibinfo {year} {2007})}\BibitemShut {NoStop}%
\bibitem [{\citenamefont {Gatti}, \citenamefont {Sottile},\ and\ \citenamefont
  {Reining}(2015)}]{PhysRevB.91.195137}%
  \BibitemOpen
  \bibfield  {author} {\bibinfo {author} {\bibfnamefont {M.}~\bibnamefont
  {Gatti}}, \bibinfo {author} {\bibfnamefont {F.}~\bibnamefont {Sottile}},\
  and\ \bibinfo {author} {\bibfnamefont {L.}~\bibnamefont {Reining}},\
  }\bibfield  {title} {\enquote {\bibinfo {title} {Electron-hole interactions
  in correlated electron materials: Optical properties of vanadium dioxide from
  first principles},}\ }\href {https://doi.org/10.1103/PhysRevB.91.195137}
  {\bibfield  {journal} {\bibinfo  {journal} {Phys. Rev. B}\ }\textbf {\bibinfo
  {volume} {91}},\ \bibinfo {pages} {195137} (\bibinfo {year}
  {2015})}\BibitemShut {NoStop}%
\bibitem [{\citenamefont {Vinson}\ and\ \citenamefont
  {Rehr}(2012)}]{PhysRevB.86.195135}%
  \BibitemOpen
  \bibfield  {author} {\bibinfo {author} {\bibfnamefont {J.}~\bibnamefont
  {Vinson}}\ and\ \bibinfo {author} {\bibfnamefont {J.~J.}\ \bibnamefont
  {Rehr}},\ }\bibfield  {title} {\enquote {\bibinfo {title} {Ab initio
  bethe-salpeter calculations of the x-ray absorption spectra of transition
  metals at the $l$-shell edges},}\ }\href
  {https://doi.org/10.1103/PhysRevB.86.195135} {\bibfield  {journal} {\bibinfo
  {journal} {Phys. Rev. B}\ }\textbf {\bibinfo {volume} {86}},\ \bibinfo
  {pages} {195135} (\bibinfo {year} {2012})}\BibitemShut {NoStop}%
\bibitem [{\citenamefont {Abbate}\ \emph {et~al.}(1991)\citenamefont {Abbate},
  \citenamefont {de~Groot}, \citenamefont {Fuggle}, \citenamefont {Ma},
  \citenamefont {Chen}, \citenamefont {Sette}, \citenamefont {Fujimori},
  \citenamefont {Ueda},\ and\ \citenamefont {Kosuge}}]{PhysRevB.43.7263}%
  \BibitemOpen
  \bibfield  {author} {\bibinfo {author} {\bibfnamefont {M.}~\bibnamefont
  {Abbate}}, \bibinfo {author} {\bibfnamefont {F.~M.~F.}\ \bibnamefont
  {de~Groot}}, \bibinfo {author} {\bibfnamefont {J.~C.}\ \bibnamefont
  {Fuggle}}, \bibinfo {author} {\bibfnamefont {Y.~J.}\ \bibnamefont {Ma}},
  \bibinfo {author} {\bibfnamefont {C.~T.}\ \bibnamefont {Chen}}, \bibinfo
  {author} {\bibfnamefont {F.}~\bibnamefont {Sette}}, \bibinfo {author}
  {\bibfnamefont {A.}~\bibnamefont {Fujimori}}, \bibinfo {author}
  {\bibfnamefont {Y.}~\bibnamefont {Ueda}},\ and\ \bibinfo {author}
  {\bibfnamefont {K.}~\bibnamefont {Kosuge}},\ }\bibfield  {title} {\enquote
  {\bibinfo {title} {Soft-x-ray-absorption studies of the electronic-structure
  changes through the ${\mathrm{vo}}_{2}$ phase transition},}\ }\href
  {https://doi.org/10.1103/PhysRevB.43.7263} {\bibfield  {journal} {\bibinfo
  {journal} {Phys. Rev. B}\ }\textbf {\bibinfo {volume} {43}},\ \bibinfo
  {pages} {7263--7266} (\bibinfo {year} {1991})}\BibitemShut {NoStop}%
\end{thebibliography}%

\end{document}